# Two-Dimensional $Na_2LiAlP_2$ Crystal for High-Performance Field-Effect Transistors

Run-Jie Peng,[1,#] Xing-Yu Wang,[1,2,#] Jun-Hui Yuan,[1,*] Nian-Nian Yu,[1] Kan-Hao Xue,[3,*] Jiafu Wang,[1] and Pan Zhang[4,*]

[1]School of Physics and Mechanics, Wuhan University of Technology, Wuhan 430070, China

[2]School of Materials and Microelectronics, Wuhan University of Technology, Wuhan 430070, China

[3]School of Integrated Circuits, Huazhong University of Science and Technology, Wuhan 430074, China

[4]School of Integrated Circuits, Peking University, Beijing 100871, China; National Key Laboratory of Advanced Micro and Nano Manufacture Technology, Beijing 100871, China

**[*]Corresponding Author**

E-mail: yuanjh90@163.com (J.-H. Yuan); xkh@hust.edu.cn (K.-H. Xue); panzhang1989@pku.edu.cn (P. Zhang)

R.-J. Peng and X.-Y. Wang contributed equally to this work.

## Abstract

High-performance, low-power transistors are core components of advanced integrated circuits, and the ultimate limitation of Moore's law has made the search for new alternative pathways an urgent priority. Two-dimensional (2D) materials have become the most promising exploration target due to their exceptional electronic properties and scalability. In this work, we conducted device transport research on the previously proposed 2D quaternary semiconductor $Na_2LiAlP_2$ using the non-equilibrium Green's function method. The results demonstrate that even with a channel length of 5 nm, $Na_2LiAlP_2$ still exhibits excellent n-type transistor characteristics, fully meeting and surpassing the technical specifications outlined in the International Roadmap for Devices and Systems (IRDS). Encouragingly, the device can easily achieve the required on-state current of 900 μA/μm under low operating voltages of 0.1 V and 0.2 V. Moreover, at 0.1 V operating voltage, the device's subthreshold swing breaks through the theoretical limit of 60 mV/dec, reaching an astonishing value 30.33 mV/dec. Additionally, its p-type transistor performance also stands out with a subthreshold swing of ~50 mV/dec when the channel length is 7 nm. Our research not only showcases the exceptional transistor properties of $Na_2LiAlP_2$ but also further expands the research scope of 2D high-performance transistors.

## 1. Introduction

Since the inception of the semiconductor industry in the mid-20th century, silicon-based transistors have dominated the development of global information technology, thanks to their mature fabrication processes and robust interfaces to their natural oxides. However, as Moore's law approaches its physical limits,[1] silicon-based technology faces three core bottlenecks: First, the quantum tunneling effect occurs when the gate length is reduced below 10 nm, where the wave-like nature of electrons allows them to traverse energy barriers, leading to a surge in leakage current.[2] For instance, in traditional silicon-based FinFETs at the 5-nm node, the leakage current density exceeds 1000 A/cm$^2$, accounting for up to 40% of static power consumption. Second, a thermal management crisis arises as silicon's thermal conductivity is merely 150 Wm$^{-1}$K$^{-1}$, causing chip temperatures to exceed 100°C under high-density integration, directly leading to device failure.[3] Finally, there is a decline in carrier mobility, as carriers in ultra-thin silicon bodies experience heavier surface scattering, significantly reducing mobility and affecting device switching speed as well as driving capability.[4]

Two-dimensional (2D) materials, with their atomic-scale thickness (0.3–1.5 nm), effectively suppress short-channel effects, enabling the realization of sub-1 nm gate-length transistors.[5] For example, monolayer molybdenum disulfide ($MoS_2$), with its direct bandgap (1.8 eV) and ultra-high electron mobility (200 cm$^2$V$^{-1}$s$^{-1}$),[6] maintains an ON/OFF ratio exceeding $10^5$ and a subthreshold swing (*SS*) as low as 60 mV/dec,[7] approaching theoretical limits, even at a 1 nm gate length.[5] Additionally, the absence of dangling bonds on the surface of 2D materials alleviates carrier scattering, leading to high theoretical carrier mobility and the potential for faster switching speeds and higher on-currents.[8–13] Therefore, transistors based on 2D materials represent a physical necessity for addressing the core scaling challenges of silicon-based technology and advancing integrated circuit technology toward smaller sizes, higher performance, and lower power consumption.[14]

However, practical applications of 2D materials face numerous limitations, such as the inverse relationship between bandgap and mobility, as well as the trade-off between surface sensitivity and stability. For instance, graphene exhibits high mobility but zero bandgap, while transition metal dichalcogenides (TMDCs) have suitable bandgaps but low mobility. On the other hand, black phosphorus offers high mobility but suffers from poor environmental stability, severely degrading device performance. Consequently, there is an urgent need to identify 2D semiconductor materials with moderate bandgaps yet high mobility. Recently, our research group proposed a new class of quaternary compound semiconductors, $A_2BXY_2$ (A = K, Na; B = Li, Na; X = Al, Ga, In; Y = P, As, Sb), which have been theoretically confirmed its experimental feasibility through the analysis of exfoliation energies.[15] This system features a stable 1D chain or 2D mesh structure composed of III-V elements ($[XY_2]^{3-}$), with suitable bandgaps (0.78–1.94 eV) and ultra-high theoretical carrier mobility ($10^4$~$10^5$ $cm^2V^{-1}s^{-1}$). Current research indicates that 2D quaternary compounds $A_2BXY_2$ exhibit significant potential in transistor applications. Therefore, in-depth studies on the transport characteristics and performance limits of MOSFETs based on the $A_2BXY_2$ system are of developmental significance for overcoming traditional physical limits, driving device miniaturization, and promoting innovations in low-power integrated circuit technology.

In this work, based on non-equilibrium Green's function (NEGF) method, we systemically study an exemplary case of $Na_2LiAlP_2$ monolayer for its transport properties. Our calculations reveal that 2D $Na_2LiAlP_2$ possesses a direct bandgap of 1.95 eV and the MOSFETs hold an excellent quantum-transport characteristic. In particular, the ON-state current of 5 nm gate 2D $Na_2LiAlP_2$ n-type MOSFETs can exceed 16000 μA/μm for high-performance (HP) applications. Furthermore, we assess the impact of low supply voltage on device performance against IRDS requirements. Finally, the performances of single device are evaluated and compared to some of the recently proposed 2D materials and its 5 nm gate length MOSFETs devices.

## 2. Methodology

The crystal structure optimization and electronic structure calculations were performed using the DS-PAW software within Device Studio provided by HZWTECH, which is based on the plane wave basis set with a cutoff energy of 600 eV.[16] The NEGF method in the Nanodcal package[17,18] was further employed to compute the carrier transport properties of the $Na_2LiAlP_2$ channels. The local density approximation was employed for the description of exchange−correlation energy.[19] The double-zeta polarized (DZP) basis set was used. To guarantee accuracy while reducing computational time, the equivalent cutoff energy for the density grid in real space was set to 80 Hartree. The self-consistent runs of the $Na_2LiAlP_2$ device in the *a*-direction and *b*-direction were carried out using 1×10×1 and 10×1×1 *k*-point grids, respectively, and their I-V curves were simulated using $k$-point grids of 1×200×1 and 200×1×1. The electrode temperature was set to 300 K. For the parameter settings of the MOSFET, we adopted the ITRS 2028 standard as the benchmark. The doping concentration was set to $1\times10^{14}$ $cm^{-2}$, a value that has been experimentally achieved in 2D materials.[20] $SiO_2$ was used as the dielectric layer, with a relative permittivity ($\varepsilon_r$) of 3.9 and a thickness of 4.1 nm. The gate length was consistent with the channel length, and the supply voltage followed the ITRS standard.

### 3. Result and discussion

**Figure 1a** illustrates the crystal structure of monolayer $Na_2LiAlP_2$. It belongs to the typical orthorhombic crystal system, with each unit cell containing 8 Na atoms, 4 Li atoms, 4 Al atoms, and 8 P atoms. Calculations reveal lattice constants of a = 11.43 Å and b = 5.70 Å, with an atomic layer thickness ($h$) of 4.74 Å. The crystal structure comprises five atomic layers, with Na metal layers on both sides and a central $[AlP_2]^{3-}$ 2D grid structure formed by Group III-V elements Al and P, where the vacancies in the grid are filled by alkali metal Li. Our previous studies have indicated that the alkali metals Li and Na in $Na_2LiAlP_2$ not only act as electron donors but also play a role in stabilizing the crystal structure, while the electronic properties are predominantly governed by the Group III-V elements Al and P.[15] Calculation results show that monolayer $Na_2LiAlP_2$ is a Γ-Γ direct gap semiconductor, with both the conduction band minimum (CBM) and valence band maximum (VBM) located at the center point of the Brillouin zone (**Figure 1b**). The bandgap calculated using the GGA-PBE functional is 1.39 eV, while that obtained with the HSE06 hybrid functional is 1.95 eV, as depicted in **Figure 1b**. Additionally, the effective mass of carriers is a crucial parameter for evaluating material transport properties. It turns out that the electron effective mass in monolayer $Na_2LiAlP_2$ is 0.11 $m_0$ along the a-direction and 0.48 $m_0$ along the b-direction; for holes, it is 0.14 $m_0$ along the a-direction and 0.43 $m_0$ along the b-direction. Our earlier computational results also indicate that the acoustic phonon-limited electron mobility of monolayer $Na_2LiAlP_2$ reaches as high as $4.90/21.1\times10^3$ $cm^2V^{-1}s^{-1}$ along the a/b directions, respectively, while the hole mobility attains $6.76/2.45\times10^3$ $cm^2V^{-1}s^{-1}$.[15] Such high carrier mobilities are key to enhancing the transistor performance.

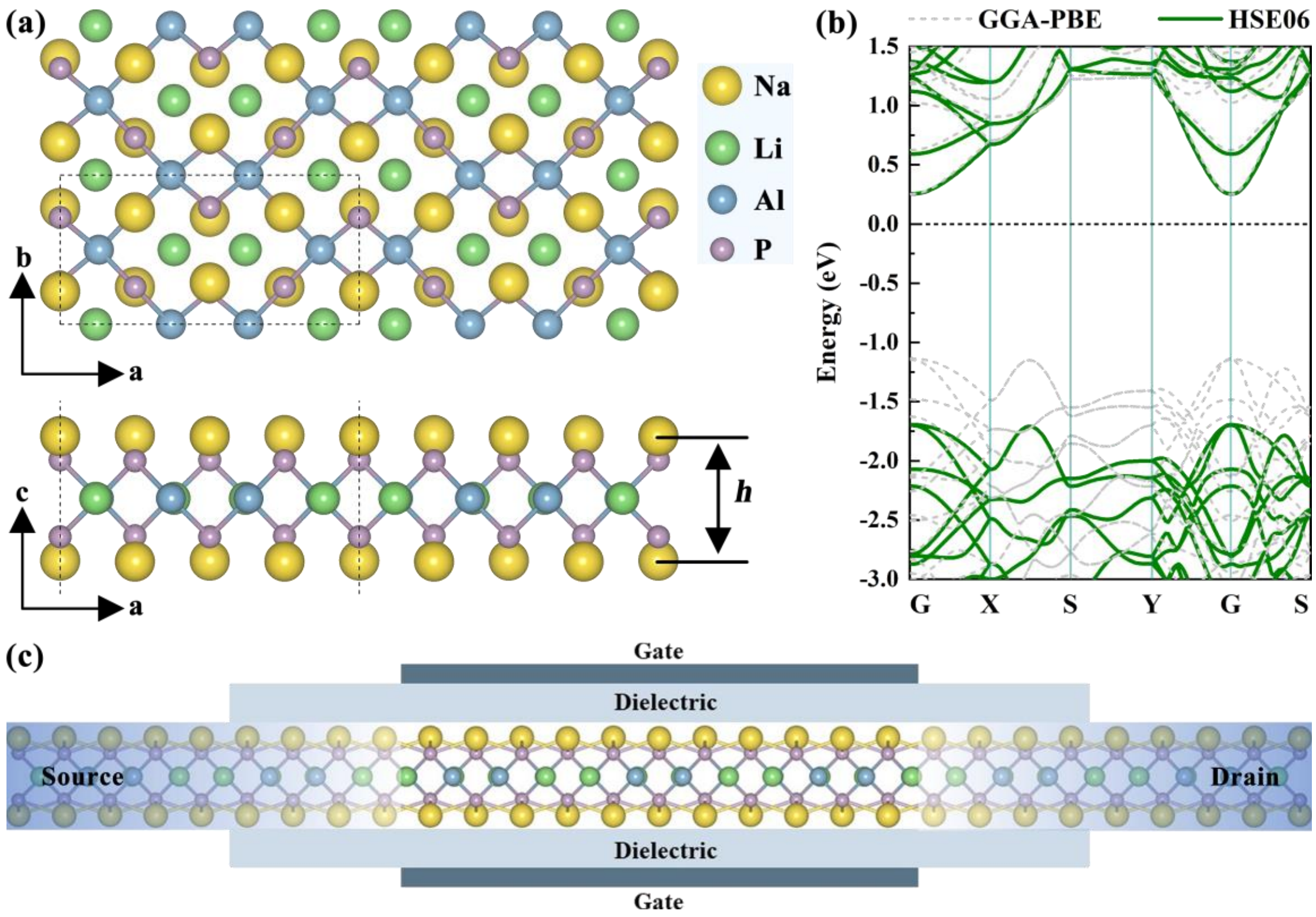

**Figure 1.** (a) Schematic of the optimized structure of monolayer $Na_2LiAlP_2$. (b) GGA-PBE and HSE06 band structure of $Na_2LiAlP_2$. (c) Schematic illustration of monolayer $Na_2LiAlP_2$ double-gated FET device.

Given that $Na_2LiAlP_2$ exhibits exceptionally superior electronic structural properties, we have further explored its potential for applications in the field of transistors. **Figure 1c** clearly presents the structure of a field-effect transistor device constructed based on $Na_2LiAlP_2$. A double-gated transistor has been considered, with a 2D $Na_2LiAlP_2$ channel embedded in $SiO_2$. Since the minimum channel length for post-Moore's law logic devices specified by IRDS standard is 12 nm, we adopted device parameters based on the IRDS standard for a 12-nm channel, such as equivalent oxide thickness (EOT) and supply voltage ($V_{\mathrm{DD}}$). For devices with channel lengths below 10 nm, the parameters are based on the ITRS standard. Since the shortest channel length of the post-Moore logic device in the IRDS standard is 12 nm, we adopt the parameters of devices, such as equivalent oxide thickness (EOT) and supply voltage ($V_{\mathrm{DD}}$), with 12-nm channel based on IRDS standard, while those below 10 nm based on ITRS standard. After extensive testing, the source and drain regions of the n-type MOSFET are optimally doped with a donor concentration of $N_{\mathrm{D}} = 1 \times 10^{14}\ cm^{-2}$ to maximize the benefits of the material's low effective electron mass. In the transfer characteristics, the ON-state current ($I_{\mathrm{ON}}$) is obtained at the ON-state voltage $V_{\mathrm{GS(ON)}} = V_{\mathrm{GS(OFF)}} + V_{\mathrm{DD}}$. Following IRDS and ITRS, $V_{\mathrm{GS(OFF)}}$ is defined by the OFF-state current ($I_{\mathrm{OFF}}$) at 0.01 and 0.1 $\mu\mathrm{A}/\mu\mathrm{m}$, respectively.

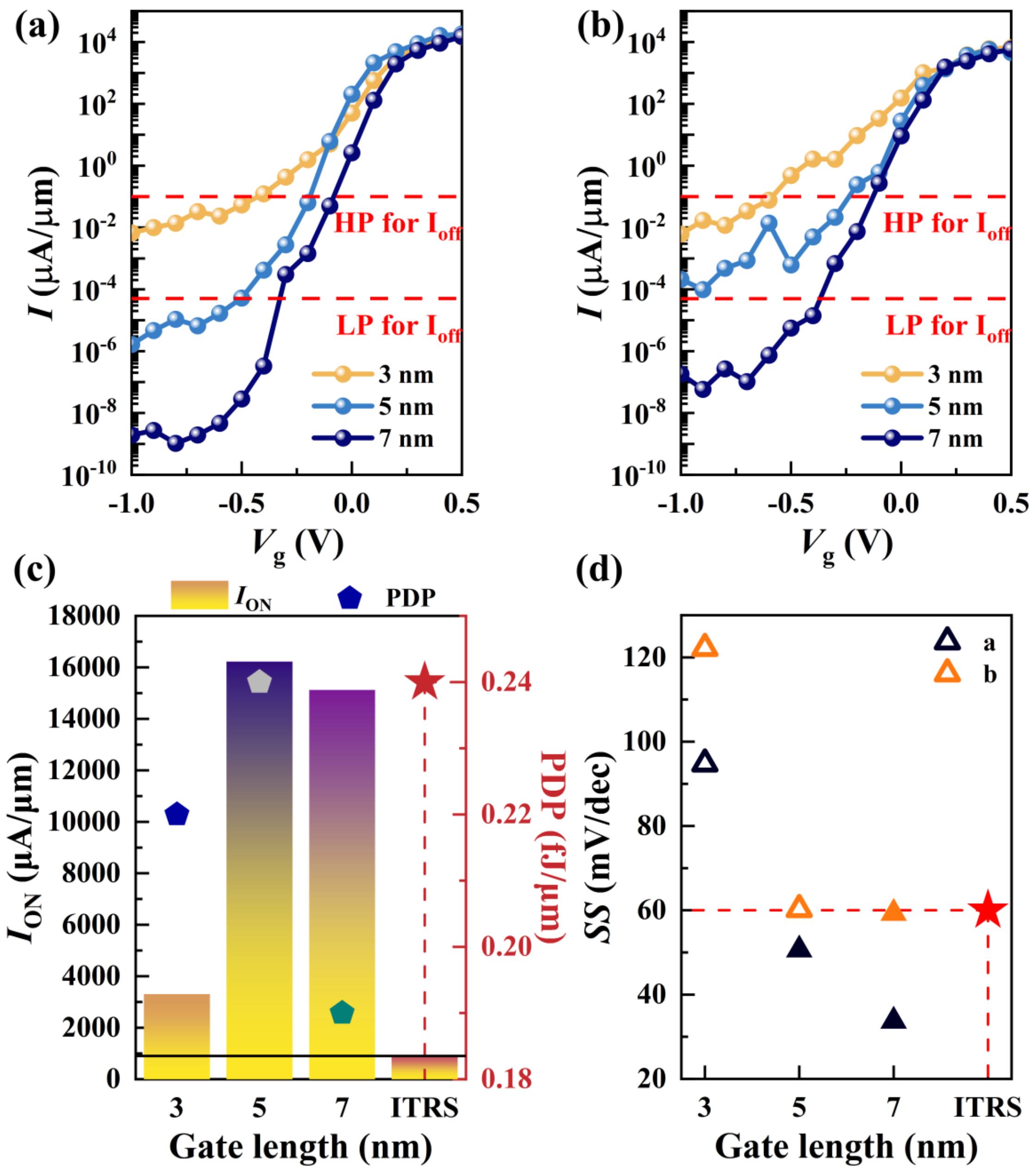

**Figure 2.** The characteristics of $Na_2LiAlP_2$ n-type MOSFETs with varying gate lengths: (a) and (b) display the transfer characteristic curves for transport along the a-direction and the b-direction, respectively; (c) shows the on-state current and power-delay product (PDP) of MOSFETs with different gate lengths when transporting along the a-direction; (d) presents the subthreshold swing of MOSFETs with different gate lengths when transporting along the a-direction.

**Figures 2a** and **2b** present the transfer characteristic curves of $Na_2LiAlP_2$ MOSFETs along two distinct directions (a and b) under varying gate lengths. Specifically, **Figure 2a** accurately depicts the transfer characteristic curves for transport along the a-direction, while **Figure 2b** meticulously illustrates those for transport along the b-direction, with all gate lengths involved being less than 12 nm. In terms of performance, all devices exhibit sufficiently large on-state currents ($I_{ON}$) that fully meet the stringent requirements of ITRS (>900 μA/μm). However, as the gate length gradually decreases, the ON-state current experiences a sharp decline due to the source-drain tunneling effect in transistors, accompanied by a rapid increase in *SS*. The trend of ON-state current variation with device gate length is clearly shown in **Figure 2c**. For devices with relatively longer gate lengths, the in-band tunneling effect is effectively suppressed, significantly improving the ON/OFF ratio and *SS*. Notably, regardless of gate length variations, the transport performance of $Na_2LiAlP_2$ MOSFETs along the a-direction consistently outperforms that along the b-direction. Specifically, when transporting along the a-direction, the device more easily reaches the OFF-state and exhibits a higher ON-state current and a lower subthreshold swing. Therefore, in **Figures 2c** and **2d**, we particularly highlight the comparisons of ON-state current, power-delay product (PDP), and *SS* for $Na_2LiAlP_2$ MOSFETs transporting along the a-direction under various gate lengths. Remarkably, a $Na_2LiAlP_2$ n-type MOSFET with a gate length of 7 nm achieves a remarkable ON-state current of $1.51\times10^4$ μA/μm, reaching an ultra-high ON-state current level (HP standard). Even more impressively, when the channel length is reduced to just 5 nm, the device's ON-state current soars to $1.62\times10^4$ μA/μm (HP), far exceeding the $0.9\times10^3$ μA/μm required by the ITRS standard.

As another critical characteristic of field-effect transistors, we conducted an in-depth evaluation of the PDP for $Na_2LiAlP_2$ MOSFETs with varying gate lengths during transport along the a-direction. It follows from **Figure 2c** that, as the gate length gradually decreases, the PDP of $Na_2LiAlP_2$ MOSFETs exhibits a trend of initially increasing and then decreasing. The PDP value marked as "ITRS" in the figure corresponds to a transistor with a gate length of 5 nm. Throughout the evaluation process, the calculation method for PDP remained consistent for $Na_2LiAlP_2$ MOSFETs with different gate lengths, and the simulation settings strictly adhered to the ITRS standards. Specific parameter settings are detailed in **Table 1** and **Table 2**. The PDP calculation formula is: PDP = $(Q_{ON}\text{-}Q_{OFF})/(V_{DD}/W)$, where $Q$ represents the total channel charge accurately calculated, and $W$ denotes the channel width. Compared with the standards set by the ITRS (0.24 fJ/μm), the $Na_2LiAlP_2$ MOSFETs exhibit lower switching energy consumption during rapid switching when the gate length is 7 nm and 3 nm, respectively. Specifically, when the gate length is 7 nm, the PDP is 0.19 fJ/μm; when the gate length is 3 nm, the PDP is 0.22 fJ/μm.

In addition, it is widely acknowledged in the industry that the *SS* serves as another crucial performance metric for evaluating the gate electrostatic control capability of logic devices. Therefore, we conducted a detailed study on the subthreshold swing of $Na_2LiAlP_2$ MOSFETs with varying gate lengths during transport along the a-direction.

When the $L_g$ is 7 nm, the $SS$ values for n-type MOSFETs along the a-direction and b-direction are 33.74 mV/dec and 59.31 mV/dec, respectively, both surpassing the Boltzmann limit (60 mV/dec). As the $L_g$ decreases, the $SS$ value increases, indicating a decline in gate control capability. Compared to MOSFETs along the b-direction, MOSFETs along the a-direction exhibit steeper current-voltage ($I$-$V$) curves and lower $SS$ values.

Furthermore, low-voltage design, as a core and pivotal technology driving the advancement of integrated circuits, plays a critical role in both portable devices and high-performance computing. Our approach employs transistors with a low operating voltage architecture. Notably, under extreme conditions where $V_{DD}$ is as low as 0.1 V or 0.2 V and the $L_g$ is merely 5 nm, $I_{ON}$ still meets the basic requirements set by the ITRS for high-performance devices in 2028. Although reducing $V_{DD}$ directly leads to a decrease in overdrive voltage, subsequently slowing down the switching speed, it is particularly noteworthy that even at the ultra-low $V_{DD}$ level of 0.2 V, relevant performance metrics still fulfill the stringent requirements of ITRS 2028. Furthermore, the $SS$ has also been significantly reduced, outperforming the $SS$ levels of traditional MOSFETs with a 7 nm gate length. The reduction in $SS$ clearly indicates that, despite the decrease in operating voltage, the gate control capability has been notably enhanced. Additionally, we conducted a comprehensive evaluation of the performance of the 5 nm device at supply voltages $V_{DD}$ of 0.1 V and 0.2 V, as shown in **Figure 3**. It follows that when $V_{DD} \leq 0.5$ V, the saturation current ($I_{sat}$) only experiences a slight decline, and the minimum $SS$ ($SS_{min}$) is as low as 30.33 mV/dec at 0.1 V $V_{DD}$.

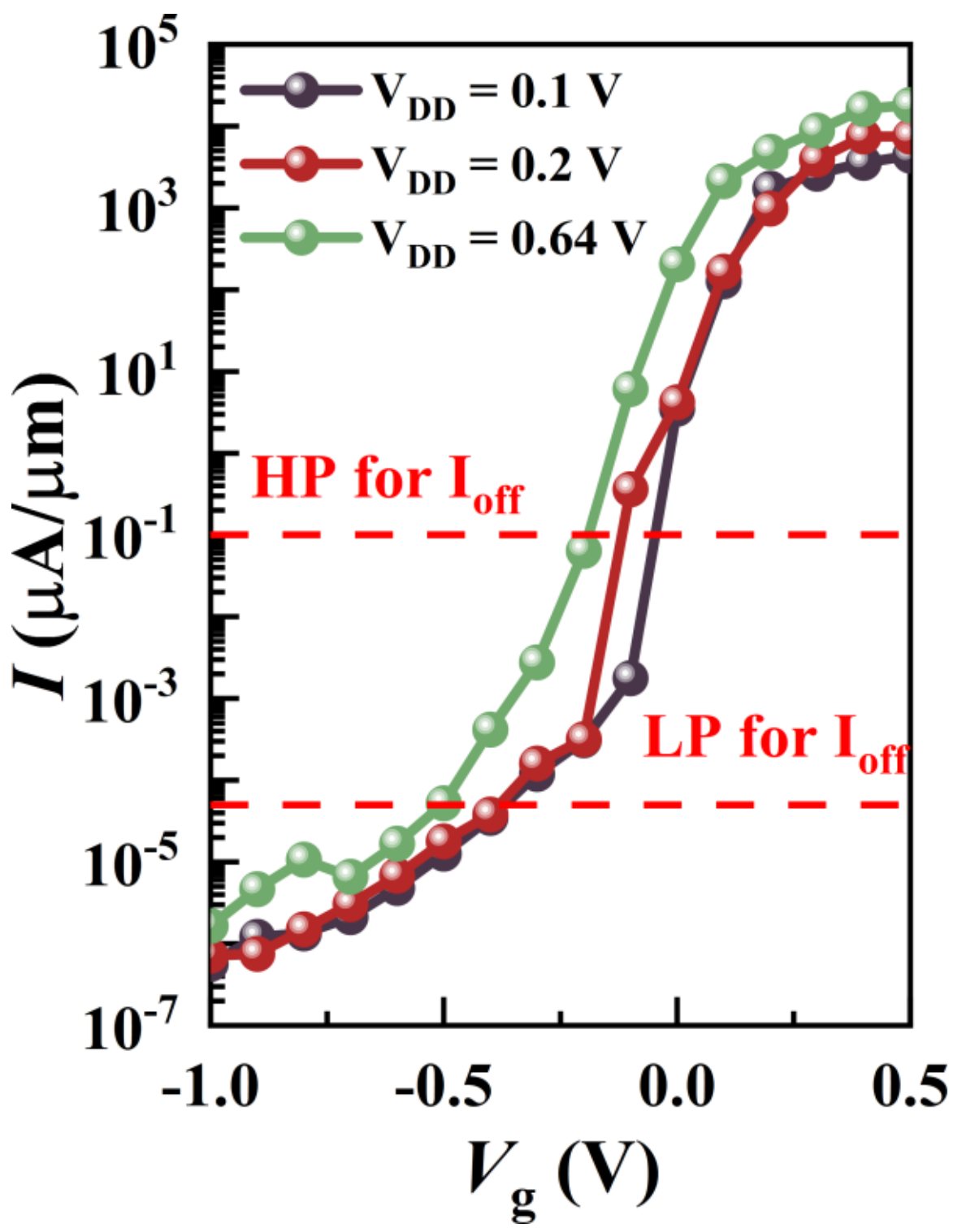

**Figure 3.** Transfer characteristics of the $Na_2LiAlP_2$ MOSFETs with different $V_{DD}$ (0.1, 0.2 and 0.64 V).

To delve deeper into the gate modulation mechanism, **Figure 4a** illustrates the local density of states (LDOS) distribution of a $Na_2LiAlP_2$ MOSFET with a 5 nm gate length, transporting along the a-direction, in its OFF-state. In the figure, the brighter yellow regions correspond to higher LDOS values. In the channel region, the barrier height (denoted as Φ) is determined by the energy barrier that carriers must overcome when traveling from the source to the drain. Therefore, within the bias window, the barrier faced by carriers from the source will not exceed this barrier height. Additionally, as the gate-source voltage ($V_{GS}$) increases, both the conduction band energy level and the barrier height in the channel region of the device decrease. Specifically, for a 5 nm $Na_2LiAlP_2$ MOSFET in terms of a-direction transport, its effective barrier height is 1.0 eV in the OFF-state under Low Power (LP) conditions. As the gate voltage increases, this effective barrier height decreases to 0.4 eV in the ON-state. Similarly, when in the OFF-state under HP conditions, its effective barrier height is 0.8 eV, and with an increase in gate voltage, it further decreases to 0.1 eV in the ON-state.

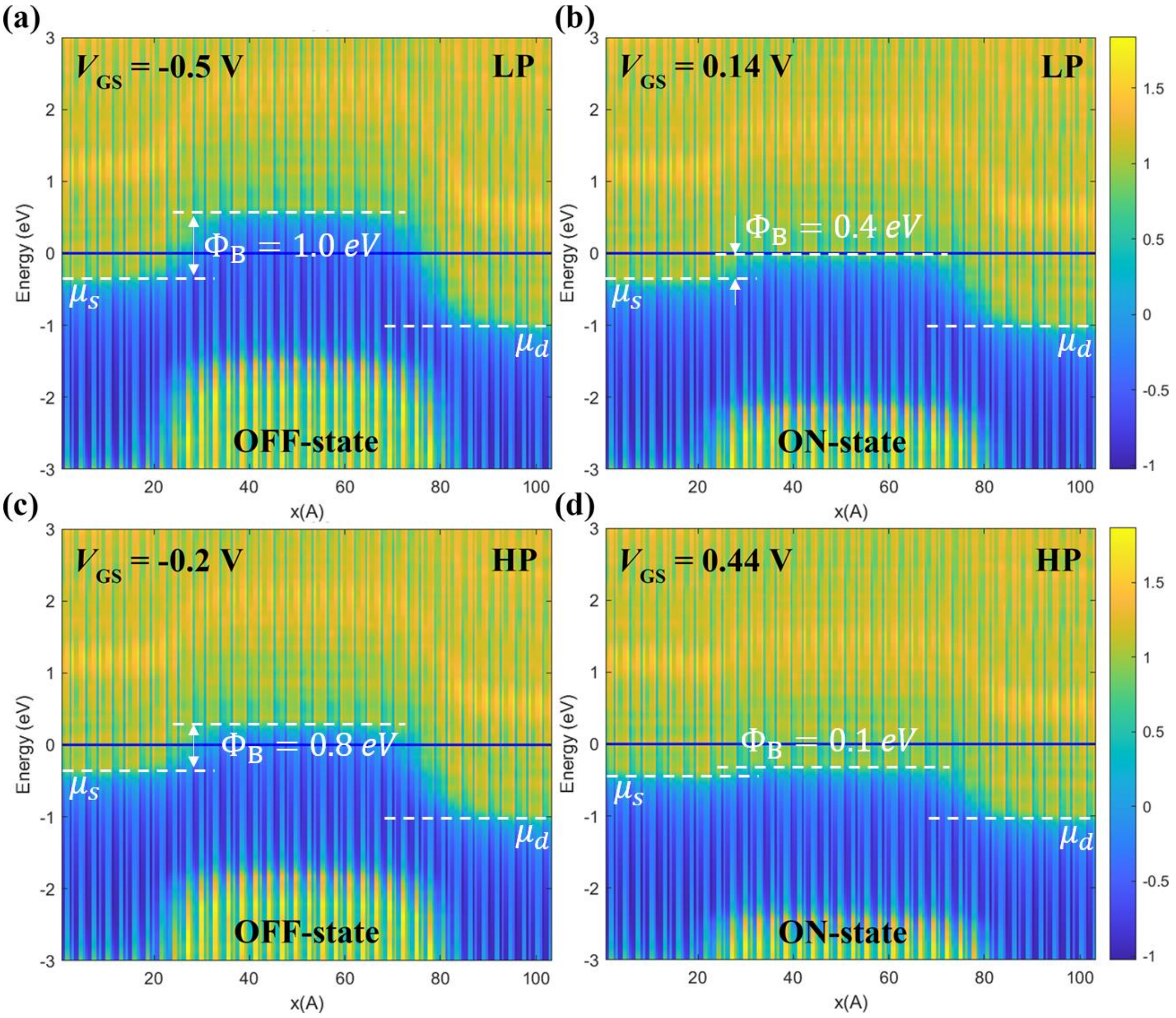

**Figure 4.** (a) Position-resolved LDOS of the 5 nm gate length $Na_2LiAlP_2$ MOSFETs at the OFF-states (HP); (b) Position-resolved LDOS of the 5 nm gate length $Na_2LiAlP_2$ MOSFETs at the ON-states (HP); (c) Position-resolved LDOS of the 5 nm gate length $Na_2LiAlP_2$ MOSFETs at the OFF-states (LP); (d) Position-resolved LDOS of the 5 nm gate length $Na_2LiAlP_2$ MOSFETs at the ON-states (LP).

**Table 1.** Benchmark of the parameters and performance metrics of HP $Na_2LiAlP_2$ MOSFETs against the ITRS requirement, including gate length $L_g$ (nm), supply voltage $V_{DD}$ (V), ON-state current $I_{ON}$ (μA/μm), ON-OFF ratio $I_{ON}/I_{OFF}$ (where $I_{ON} = 5\times10^{-5}$ μA/μm), τ (ps), PDP (fJ/μm) and subthreshold slope *SS* (mV/dec).

| Type | Direction | $L_g$ | $V_{DD}$ | $I_{ON}$ | $I_{ON}/I_{OFF}$ | $\tau$ | PDP | *SS* |
|---|---|---|---|---|---|---|---|---|
| n | a | 3 | 0.64 | 3307 | $3.307\times10^4$ | 0.10 | 0.22 | 94.78 |
| | | 5 | 0.1 | 3972 | $3.972\times10^4$ | 0.59 | 0.23 | 30.33 |
| | | | 0.2 | 7449 | $7.449\times10^4$ | 0.21 | 0.32 | 32.73 |
| | | | 0.64 | 16220 | $1.622\times10^5$ | 0.04 | 0.24 | 50.46 |
| | | 7 | 0.64 | 15127 | $1.513\times10^5$ | 0.02 | 0.19 | 33.74 |
| | b | 3 | 0.64 | 1017 | $1.017\times10^4$ | 1.32 | 0.86 | 122.19 |
| | | 5 | | 5708 | $5.708\times10^4$ | 0.34 | 1.23 | 60.24 |
| | | 7 | | 5899 | $5.899\times10^4$ | 0.61 | 2.29 | 59.31 |
| p | a | 7 | 0.64 | 7034 | $7.034\times10^4$ | 0.51 | 2.62 | 32.81 |
| | | 5 | | 62.15 | $6.215\times10^2$ | 4.03 | 0.16 | 58.43 |
| ITRS 2028 | -- | 5 | 0.64 | 900 | $9.000\times10^3$ | 0.423 | 0.24 | 60 |

**Table 2.** Benchmark of the parameters and performance metrics of LP $Na_2LiAlP_2$ MOSFETs against the ITRS requirement, including gate length $L_g$ (nm), supply voltage $V_{DD}$ (V), ON-state current $I_{ON}$ (μA/μm), ON-OFF ratio $I_{ON}/I_{OFF}$ (where $I_{ON} = 5\times10^{-5}$ μA/μm), τ (ps), PDP (fJ/μm) and subthreshold slope *SS* (mV/dec).

| Type | Direction | $L_g$ | $V_{DD}$ | $I_{ON}$ | $I_{ON}/I_{OFF}$ | $\tau$ | PDP | *SS* |
|---|---|---|---|---|---|---|---|---|
| n | a | 5 | 0.1 | 2624 | $5.248\times10^7$ | 0.50 | 0.13 | 30.33 |
| | | | 0.2 | 3919 | $7.838\times10^7$ | 0.25 | 0.19 | 32.73 |
| | | | 0.64 | 2121 | $4.242\times10^7$ | 0.18 | 0.25 | 50.46 |
| | | 7 | 0.64 | 5391 | $1.078\times10^8$ | 0.05 | 0.17 | 33.74 |
| | b | 7 | 0.64 | 1547 | $3.094\times10^7$ | 2.14 | 2.12 | 59.31 |
| p | a | 7 | 0.64 | 3307 | $6.614\times10^7$ | 0.82 | 1.9 | 32.81 |
| ITRS 2028 | -- | 5 | 0.64 | 295 | $5.900\times10^6$ | 1.493 | 0.28 | 60 |

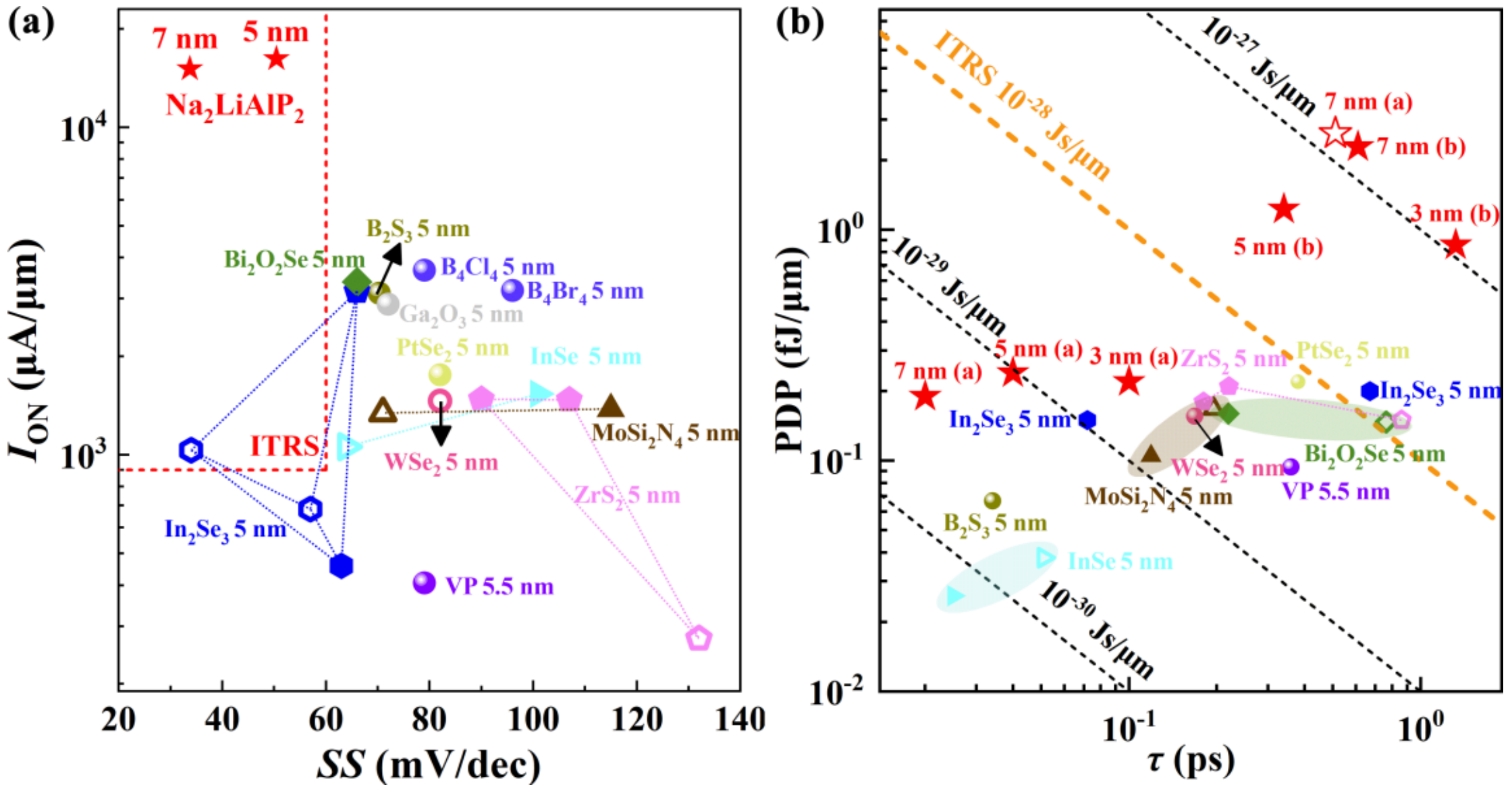

**Figure 5.** (a) ON-state current ($I_{ON}$) versus subthreshold slope (*SS*) of $Na_2LiAlP_2$ and other 2D MOSFETs with 5-nm gate length. (b) The PDP against τ of the p-type (hollow pentagram) and n-type (solid pentagram) ($L_g$ = 3, 5 and 7 nm) MOSFETs for the HP utilization, via several 2D channel materials. All n-type MOSFETs have a gate length of 5 nm, with all n-type MOSFETs marked by solid icons and all p-type MOSFETs marked by hollow icons.

Furthermore, although the performance of $Na_2LiAlP_2$ p-type MOSFETs generally lags behind that of their n-type counterparts, when the gate length is 7 nm and transport direction is along a, they still meet the requirements for high-performance devices. Moreover, their *SS* value is comparable to that of the 7 nm n-type devices, measuring 32.81 mV/dec. The aforementioned results indicate that $Na_2LiAlP_2$ can simultaneously fulfill the requirements for high-performance sub-10-nm n-type or p-type MOSFETs, promising its potential for low-dimensional electronic devices.

In addition, to achieve a clearer and more intuitive evaluation of the performance of $Na_2LiAlP_2$ MOSFETs with a 5 nm gate length transporting along the a-direction, we extracted the results of their ON-state current and *SS*. Subsequently, we conducted a comparative analysis with several 2D materials reported in recent studies, including $B_4X_4$ (X=Cl, Br),[21] phosphorene,[22] InSe,[23] and $Bi_2O_2Se$,[24] $B_2S_3$,[25] $In_2Se_3$,[26] $Ga_2O_3$,[27] $PtSe_2$,[28] violet phosphorene (VP),[29] $ZrS_2$,[30] $WSe_2$.[31] As depicted in **Figure 5a**, the $Na_2LiAlP_2$ MOSFET with a 5 nm gate length outperforms the aforementioned four materials in terms of both ON-state current and *SS*.

Last but certainly not least, the energy effectiveness of FET devices is described by the energy-delay product (EDP) by taking into account both $\tau$ and PDP concerning the association $\mathrm{EDP} = \tau \times \mathrm{PDP}$. For the semiconducting FETs having a gate length of 5nm displaying a 2D feature such as $MoSi_2N_4$,[32] $B_2S_3$,[25] $Bi_2O_2Se$,[24] $In_2Se_3$,[26] InSe,[23] $PtSe_2$,[28] violet phosphorene (VP),[29] $ZrS_2$,[30] $WSe_2$,[31] the corresponding EDPs are depicted in **Figure 5b** for comparison. It should be emphasized that the red pentagram in **Figure 5b** represents the results obtained in this study, demonstrating the potential as lower EDP-associated devices. Therefore, benchmarked against the ITRS, for the requirements of HP applications, the EDPs of 3, 5, and 7 nm n-type $Na_2LiAlP_2$ MOSFETs are $2.2\times10^{-29}$, $9.6\times10^{-30}$, and $3.8\times10^{-30}$ Js/μm, respectively, successfully meeting the ITRS requirements ($10^{-28}$ Js/μm). All these results indicate that $Na_2LiAlP_2$ is highly competitive in terms of high performance field-effect transistors. In addition, in our device design, the impact brought by the introduction of underlap structure did not take into account. In fact, previous studies have confirmed that introducing the underlap structure can effectively further mitigate the short-channel effect, thereby significantly enhancing the device performance of $Na_2LiAlP_2$ MOSFET. Coupled with the mechanically exfoliatable nature of bulk $Na_2LiAlP_2$, the performance of experimentally fabricated transistors based on this material is highly promising and worth anticipating.

**4. Conclusion**

In summary, we evaluate the device performance limit of sub-10-nm 2D $Na_2LiAlP_2$ MOSFETs by employing DFT and NEGF simulations. It is revealed that devices transporting along the a-direction exhibited significantly superior performance: under a 5 nm gate length, the subthreshold swing dropped as low as 50.46 mV/dec, successfully surpassing the Boltzmann limit; concurrently, the ON-state current reached a remarkable 16,220 μA/μm, far exceeding the standards set by the ITRS for high performance device. Notably, even under a 7 nm gate length, the subthreshold swing of the device could be further reduced to 33.74 mV/dec. Subsequently, we conducted an in-depth analysis of an n-type 5 nm gate length $Na_2LiAlP_2$ MOSFET transporting along the a-direction under low operating voltages. The results showed that the device could easily meet the IRTS requirement of 900 μA/μm for the on-state current of high-performance devices at both 0.1 V and 0.2 V operating voltages. Particularly outstanding was the achievement of an astonishingly low subthreshold swing of 30.33 mV/dec at a 0.1 V operating voltage. Additionally, we performed calculations for a p-type 7 nm gate length device transporting along the a-direction and found that it not only fully complied with the ITRS requirements but also exhibited an impressively low subthreshold swing of 32.81 mV/dec, demonstrating equally remarkable performance. Therefore, our work has not only fully demonstrated the immense application potential of the $Na_2LiAlP_2$ crystal in the field of low-dimensional transistors but also confirms the idea of enabling Moore's law down to 5-nm through 2D materials.

**ASSOCIATED CONTENT**

**AUTHOR INFORMATION**

**Corresponding Authors**

Jun-Hui Yuan − School of Physics and Mechanics, Wuhan University of Technology, Wuhan 430070, China; orcid.org/0000-0002-3892-604X; Email: yuanjh90@163.com

Kan-Hao Xue − School of Integrated Circuits, Huazhong University of Science and Technology, Wuhan 430074, Hubei, China; orcid.org/0000-0001-5195-1867; Email: xkh@hust.edu.cn

Pan Zhang − School of Integrated Circuits, Peking University, Beijing 100871, China; National Key Laboratory of Advanced Micro and Nano Manufacture Technology, Beijing, 100871, China; orcid.org/0000-0001-9130-5982; Email: panzhang1989@pku.edu.cn

**Authors**

Run-Jie Peng − School of Physics and Mechanics, Wuhan University of Technology, Wuhan 430070, China;

Xing-Yu Wang − School of Physics and Mechanics, School of Materials and Microelectronics, Wuhan University of Technology, Wuhan 430070, China;

Nian-Nian Yu − School of Physics and Mechanics, Wuhan University of Technology, Wuhan 430070, China;

Jiafu Wang − School of Physics and Mechanics, Wuhan University of Technology, Wuhan 430070, China.

**Author Contributions**

R. J. P. and X. Y. W. performed the calculation and data analysis, and wrote the original manuscript. N. N. Y. performed the data analysis. J.-H. Y. conceived the idea, performed the calculation and data analysis, and wrote and revised the manuscript. K.-H. X. conducted guidance and wrote and revised the manuscript. J. W. and P. Z. conducted guidance and manuscript revision. All authors reviewed this manuscript.

**Notes**

The authors declare no competing financial interest.

## ACKNOWLEDGMENTS

This work was partially supported the Fundamental Research Funds for the Central Universities (WUT: 2024IVA052). We gratefully acknowledge HZWTECH for providing computation facilities.